\documentstyle[12pt,epsfig]{article}             
\newcommand{\beq}{\begin{equation}}             
\newcommand{\eeq}{\end{equation}}               
\newcommand{\bqry}{\begin{eqnarray}}            
\newcommand{\eqry}{\end{eqnarray}}              
\newcommand{\bqryn}{\begin{eqnarray*}}          
\newcommand{\eqryn}{\end{eqnarray*}}            
\newcommand{\preprint}[1]{\begin{table}[t]      
            \begin{flushright}                  
            \begin{large}{#1}\end{large}        
            \end{flushright}                    
            \end{table}}                        
\newcommand{\PD}[2]                             
    {\frac{\partial^{#2}}{\partial #1^{#2}}}    
\begin{document} 
\preprint{LA-UR-02-3291} 
\title{An Analytic Model of the Shear Modulus \\ at All Temperatures 
and Densities} 
\author{\\ Leonid Burakovsky,\thanks{E-mail: BURAKOV@LANL.GOV} \
Carl W. Greeff,\thanks{E-mail: GREEFF@LANL.GOV} \ and \ 
Dean L. Preston\thanks{E-mail: DEAN@LANL.GOV} 
 \\  \\  
Los Alamos National Laboratory \\ Los Alamos, NM 87545, USA }
\date{ }
\maketitle
\begin{abstract}
\vspace*{0.1cm}
\hspace*{-0.76cm}
An analytic model of the shear modulus applicable at temperatures up to 
melt and at all densities is presented. It is based in part on a relation 
between the melting temperature and the shear modulus at melt. Experimental 
data on argon are shown to agree with this relation to within 1\%. The
model of the shear modulus involves seven parameters, all of which can 
be determined from zero-pressure experimental data. We obtain the values 
of these parameters for 11 elemental solids. Both the experimental data on 
the room-temperature shear modulus of argon to compressions of $\sim 2.5,$ 
and theoretical calculations of the zero-temperature shear modulus of 
aluminum to compressions of $\sim 3.5$ are in good agreement with the model. 
Electronic structure calculations of the shear moduli of copper and gold 
to compressions of 2, performed by us, agree with the model to within 
uncertainties. 
\end{abstract}
\bigskip

\section{Introduction} 

A reliable model of the adiabatic (isentropic) shear modulus, $G,$ of a 
polycrystalline solid at temperatures to $T_m,$ the melting temperature, 
and up to megabar pressures is needed for many applications, including the 
modeling of plastic deformation at extremes of pressure and temperature, 
numerical calculations of elastic and shock wave propagation, and even 
calculations of the oscillations of low-mass astrophysical objects.

Adiabatic elastic properties are generally determined by ultrasonic 
wave-speed measurements, which are usually made in the low-pressure regime.
Zero-pressure experimental data have been accumulated on single-crystal 
elastic constants, together with polycrystalline averages, at temperatures 
from $T=0$ to nearly $T_m$ for Ag (to within $60\,^\circ $K of $T_m$) 
\cite{Ag}, Au (to within $60\,^\circ $K of $T_m$) \cite{Au}, Ge (to within 
$90\,^\circ $K of $T_m$) \cite{Ge}, and V (to within $80\,^\circ $K of $T_m$) 
\cite{V}. The data run from $T=0$ to $T_m$ for Al \cite{Al}, Ar \cite{Ar}, 
Bi \cite{BiSn}, Cd \cite{Cd}, Cs \cite{Cs}, Cu \cite{Cu}, In \cite{InPb}, 
K \cite{K}, Na \cite{Na}, Nb \cite{Nb}, Ne \cite{Ne}, Pb \cite{InPb}, Sn 
\cite{BiSn}, Ta \cite{Ta}, Te \cite{Te}, Xe \cite{Xe}, and Zn \cite{Zn}. 

On the theoretical side, it is possible to calculate singe-crystal 
elastic constants as a function of compression at zero temperature from 
electronic-structure theory. Such calculations were done by Straub {\it et 
al.}~\cite{Straub} for Cu, Christensen {\it et al.}~\cite{CRR} for Mo and 
W, S\"{o}derlind {\it et al.}~\cite{SMW} for Fe, and S\"{o}derlind and 
Moriarty \cite{SM} for Ta. With known interatomic potentials it is 
possible to calculate the temperature dependence of the elastic constants 
by computer simulation techniques, as demonstrated by the calculations for 
Na \cite{SW}, Mg \cite{GM}, and Cu \cite{RJGA}. Bounds on the shear modulus, 
$G$, can be calculated from the single-crystal elastic constants for any 
crystal class \cite{bounds}, and for a cubic crystal the polycrystalline 
shear modulus can be calculated exactly using the Kr\"{o}ner cubic equation 
\cite{Kroner}.

Guinan and Steinberg \cite{GS} modeled the zero-temperature shear modulus 
as $G=G_0+G'_0\,P\,(\rho _0/\rho )^{1/3},$ where $G'_0$ is the pressure 
derivative of $G$ at zero pressure and $\rho $ is density. This functional 
form was chosen so that $G\sim \rho ^{\;\!4/3}$ as $\rho \rightarrow \infty ,$ 
the correct asymptotic behavior albeit with a prefactor which does not 
generally coincide with that given by the one-component plasma model for $G.$ 
Preston and Wallace \cite{PW} proposed a model for the temperature dependence 
of the shear modulus at any density, but left the density dependence itself 
arbitrary. The dependence of the shear modulus on both density and 
temperature has also been discussed by Anderson \cite{And}. 

In this paper we develop an analytic model for the density and temperature 
dependence of the shear modulus by combining four key elements. First is 
a simple but accurate relation between the density, the melting temperature 
as a function of density, $T_m(\rho),$ and the shear modulus along the 
solidus \cite{BP1,BPS}. Second is the Preston-Wallace model for the shear 
modulus. Third is an analytic model for the Gruneisen parameter \cite{BP3} 
that is used in conjunction with the fourth ingredient, the Lindemann 
criterion \cite{Young}, to generate an analytic expression for $T_m(\rho ).$ 

\section{A relation between shear modulus and melting temperature}

The melting temperature and shear modulus along the solidus approximately 
satisfy the relation
\beq
\frac{G(\rho ,T_m(\rho ))}{\rho \;T_m(\rho )}=\frac{G(\rho _{{\rm ref}},T_m(
\rho_{{\rm ref}}))}{\rho _{{\rm ref}}\;\!T_m(\rho _{{\rm ref}})},
\eeq
where $\rho _{{\rm ref}}$ is a reference density. This relation is the 
foundation of our model for the shear modulus, so we provide theoretical 
justification for it following two approaches: the theory of 
dislocation-mediated melting \cite{BP1,BPS}, 
and the theory of a Debye solid (in which it derives as a consequence 
of the proportionality of $G$ to the square of the Debye temperature). 
The relation is shown to agree very well with shear modulus data 
on argon, the only data available for such a comparison.

\subsection{Two derivations of relation (1)}

It follows from our model of melting as a dislocation-mediated phase 
transition that the relation
\beq 
k_BT_m=\frac{1-\nu (T_m)/2}{1-\nu (T_m)}\;\frac{ G(T_m)v(T_m)}{
\ln (z-1)}\;\frac{\lambda}{8\,\pi}
\ln \left( \frac{\alpha ^2}{4b^2d(T_m)}\right) . 
\eeq 
holds at any pressure. Here $b$ is the magnitude of the Burgers vector, 
$\nu $ is the Poisson ratio, $v$ is the Wigner-Seitz volume, $\lambda \equiv 
b^3/v$ is a geometric factor characterizing the lattice, $z$ is the 
coordination number, and $d(T_m)$ is the dislocation density at melt. Note 
that the factors $\lambda $ and $\ln (z-1)$ explicitly account for the 
influence of crystal structure on melting. The value of $\lambda $ is $3\sqrt{
3}/4\approx 1.30$ for body-centered cubic (bcc), and $\sqrt{2}\approx 1.41$ 
for face-centered cubic (fcc) and ideal $(c/a=\sqrt{8/3})$ hexagonal 
close-packed (hcp) lattices \cite{BPS}. The parameter $\alpha $ is the ratio 
of $b$ to the dislocation core radius, $r_0;$ $\alpha \approx 2.9$ for both 
bcc and fcc crystals \cite{BPS}. This melting relation plus experimental data 
on over half the elements in the periodic table give $b^2d(T_m)=0.61\pm 0.20$ 
(throughout this paper the error in such expressions is the corresponding 
standard deviation) with $G(300\,^\circ $K), $v_{WS}(300\,^\circ $K) used 
instead of $G(T_m),$ $v_{WS}(T_m)$, respectively \cite{BP1}.

\vspace*{0.5cm}
\begin{center}
{\footnotesize
\begin{tabular}{|l|l|l|l|l|l|l|l|l|l|l|l|}
\hline 
 element & Ba & Cs & Cr & $\delta $-Fe & K & Li & Na & Nb & Rb & $\beta $-Ti &
 V   \\
\hline 
 $T_m,$ $^\circ $K & 1000 & 301.6 & 2130 & 1811 & 336.5 & 453.7 & 370.9 & 
 2750 & 312.5 & 1941 & 2183   \\
\hline 
 $v(T_m),$ $\stackrel{\circ }{{\rm A}}$$^3$ & 66.68 & 116.8 & 13.10 & 12.76 & 
 76.38 & 22.14 & 40.17 & 19.33 & 93.37 & 18.61 & 14.89   \\
\hline 
 $G(T_m),$ GPa & 2.96 & 0.39 & 35.7 & 30.8 & 0.80 & 3.60 & 1.93 & 32.6 & 0.60 
 & 21.9 & 32.3   \\
\hline 
 $G\,v/(k_BT_m)$ & 14.3 & 10.9 & 15.9 & 15.7 & 13.2 & 12.7 & 
 15.1 & 16.6 & 13.0 & 15.2 & 15.9   \\
\hline 
\end{tabular}
}
\end{center}
\vspace*{0.1cm}
Table 1. Numerical values of the ratio $G(T_m)v(T_m)/(k_BT_m)$ for 11 
elemental solids that melt from bcc crystalline structure at normal pressure. 
 \\

\begin{center}
{\footnotesize
\begin{tabular}{|l|l|l|l|l|l|l|l|l|l|l|l|}
\hline 
 element & Ag & Al & Ar & Au & $\beta $-Co & Cu & Ni & Pb & Pd & Pt & Rh   \\
\hline 
 $T_m,$ $^\circ $K & 1235 & 933.5 & 83.8 & 1338 & 1768 & 1358 & 1728 & 
 600.6 & 1828 & 2041 & 2237   \\
\hline 
 $v(T_m),$ $\stackrel{\circ }{{\rm A}}$$^3$ & 18.19 & 17.55 & 40.90 & 17.88 & 
 11.96 & 12.61 & 11.85 & 31.14 & 15.65 & 16.04 & 14.87   \\
\hline 
 $G(T_m),$ GPa & 17.2 & 15.6 & 0.60 & 15.2 & 34.7 & 27.1 & 38.6 & 5.60 & 35.0 
 & 32.0 & 55.0   \\
\hline 
 $G\,v/(k_BT_m)$ & 18.4 & 21.2 & 21.2 & 14.7 & 17.0 & 18.2 & 19.2 & 21.0 & 
 21.7 & 18.2 & 26.5   \\
\hline 
\end{tabular}
}
\end{center}
\vspace*{0.1cm}
Table 2. Numerical values of the ratio $G(T_m)v(T_m)/(k_BT_m)$ for 11 
elemental solids that melt from fcc crystalline structure at normal pressure. 
 \\

From the compilation of data in Tables 1 and 2 we find that the product of 
$\lambda $ and the logarithm in Eq.\ (2) (with $\nu (T_m)=0.42\pm 0.02$ 
\cite{Bel}) is a constant to 15\% at zero pressure: 
\beq
\frac{\lambda }{8\pi }\;\!\ln \left( \frac{\alpha ^2}{4b^2d(T_m)}\right) = 
\left[
\begin{array}{ll}
0.100\pm 0.015, & {\rm bcc,} \\
0.091\pm 0.014, & {\rm fcc.}
\end{array}
\right.
\eeq

We make the reasonable assumption that the mean interdislocation distance at 
the melting point, $2R\approx 1/\sqrt{d(T_m)}$, scales with $b$, which implies 
that $b^2d(T_m)$ is a compression-independent constant. It is also assumed 
that $\alpha ^{-1}=r_0/b$ is unchanged under compression. Hence $\lambda \ln 
(\alpha ^2/4b^2d)$ is expected to be pressure-independent with approximately 
the same value for both bcc and fcc elements. It then follows from (2) that 
for a given element
\beq
\xi (P)\equiv \frac{1-\nu (P,T_m(P))/2}{1-\nu (P,T_m(P))}\;
\frac{G(P,T_m(P))v(P,T_m(P))}{k_BT_m(P)\ln (z-1)}=c,
\eeq 
where the constant $c$ has nearly the same value for both bcc and fcc 
elements. Experimental validation of this relation is not posible because of 
a lack of data from moderate to high compressions. However, the $P\rightarrow 
0$ and $P \rightarrow \infty $ limits are consistent with Eq.~(4), which we 
now demonstrate. 

At very high compressions 
a solid becomes a crystallized one-component plasma (OCP), i.e., a lattice 
of ions in a uniform neutralizing background of electrons \cite{Young}. The 
melting curve of a solid at ultrahigh pressures is described by the equation
\beq
\frac{Z^2e^2}{a(T_m)k_BT_m}=\Gamma _m,
\eeq
where $Z$ is the atomic number, $a=(3v/4\pi )^{1/3}$ is the Wigner-Seitz 
radius, and $\Gamma _m$, a dimensionless constant, is the OCP coupling 
parameter at melt \cite{Young}. Numerous calculations of $\Gamma _m$ for 
a bcc OCP crystal (see ref.\ \cite{BP2} for a review) converge on the value 
175 \cite{DS,PC}. The value of $\Gamma _m$ for a fcc OCP crystal has been 
calculated to be $196\pm 1$ \cite{HCV} and 208.3 \cite{Ree}; hence we take 
$\Gamma _m=200$ for a fcc OCP crystal in the following analysis. The bcc OCP 
single-crystal elastic constants $(c_{11}-c_{12})/2$ and $c_{44}$ have been 
calculated by means of Monte-Carlo simulations \cite{OI}. A linear fit to 
the values of $G$ given by the formula of Sisodia {\it et al.} \cite{SDV} 
(when $c_{11}$ and $c_{12}$ are not known separately, the value of $G$ 
given by this formula approximates Kr\"{o}ner's shear modulus with high 
accuracy and, in fact, tends to the precise Kr\"{o}ner value in the limit 
$P\rightarrow \infty )$ results in \cite{BP2} 
\beq
G_{{\rm bcc}}^{{\rm OCP}}(T)=g_{{\rm bcc}}\,\left ( \frac{4\pi }{3}\right) ^{
1/3}\frac{Z^2e^2}{v^{4/3}}\left( 1-\beta ^{{\rm OCP}}_{{\rm bcc}}\;\!\frac{
T}{T_m}\right),
\eeq
where $g_{{\rm bcc}}=0.09301$ and $\beta ^{\rm OCP}_{{\rm bcc}}=0.21\pm 0.18.$ 
We have calculated (unpublished) the coefficient $g_{{\rm fcc}}$ to be 
0.09011. The coefficient $\beta ^{{\rm OCP}}_{{\rm fcc}}$ has not been 
calculated, so we assume $\beta ^{{\rm OCP}}_{{\rm fcc}}=\beta ^{{\rm OCP}}_{
{\rm bcc}}.$ We have also calculated the Voigt (V) and Reuss (R) bounds on the 
shear modulus of an ideal hcp OCP crystal: $g_{{\rm hcp}}^{\rm V}=0.1194,$ 
$g_{{\rm hcp}}^{\rm R}=0.1045,$ hence $g_{{\rm hcp}}=0.1120$ for the 
Voigt-Reuss-Hill average.  

From Eqs.\ (4) and (6), and the ultrahigh pressure limit $\nu (T)=1/2$ 
\cite{Kop,BPS2}, we obtain $\xi ^{{\rm OCP}}_{{\rm bcc}}=9.9\pm 2.3$ and 
$\xi ^{{\rm OCP}}_{{\rm fcc}}=8.9\pm 2.0.$ Comparison of the OCP values of 
$\xi $ to their zero-pressure counterparts (which follow from Eqs.\ (2) and 
(3)), $\xi _{{\rm bcc}}(0)=10.0\pm 1.5$ and $\xi _{{\rm fcc}}(0)=11.0\pm 1.7,$ 
shows that the $P=0$ and OCP values agree to within uncertainties, compelling 
evidence, though not a proof, that Eq.\ (4) is in fact valid, at least for 
bcc and fcc lattices. The uncertainty-weighted average of the four values is 
$10.0\pm 1.8$. 

Formula (1) now follows from Eq.\ (4) provided that the ratio $(1-\nu (T_m)/
2)/(1-\nu (T_m))$ is (approximately) a constant; in fact this ratio varies 
between $\approx 4/3$ at $P=0$ and $3/2$ as $P\rightarrow \infty ,$ i.e., it 
is $(17\pm 1)/12\approx 17/12$ to 94\% accuracy. 

Formula (1) can also be derived from the theory of a Debye solid. 
Ledbetter \cite{Ledb} derived the Debye-solid relation 
\beq
\Theta _D=\frac{\Lambda }{v^{1/3}}\;\!\sqrt{\frac{G}{\rho }},
\eeq
where $\Theta _D$ is the Debye temperature and $\Lambda $ is a constant. 
(Since $G\sim \rho ^{\;\!4/3}$ as $\rho \rightarrow \infty ,$ $\Theta_D\sim 
\rho ^{\;\!1/2},$ which is consistent with $\gamma $ (Gr\"{u}neisen) 
$\rightarrow 1/2$ \cite{BP3,Kop}. Its widely used counterpart \cite{Ledb}, 
$\Theta _D=\tilde{\Lambda }v^{-1/3}\;\!\sqrt{B/\rho }$, where $B$ is the bulk 
modulus, has the wrong asymptotic behavior, $\Theta _D\sim \rho ^{\;\!2/3}.$) 
Siethoff and Ahlborn \cite{SA} demonstrated the validity of the Ledbetter 
formula at $P=0$ for Debye-like cubic solids \cite{SA,S1,S2}, non-Debye 
hexagonal and tetragonal solids \cite{S3}, and intermetallic compounds 
\cite{S4}. Eq.\ (7), $v\sim 1/\rho ,$ and the Lindemann melting criterion 
\cite{Young} 
\beq
\frac{T_m(\rho )\;\!\rho ^{\;\!2/3}}{\Theta _D^2(\rho )}={\rm constant,}
\eeq  
again yield relation (1). 

\vspace*{0.5cm}
\begin{center}
\begin{tabular}{|r|c|c|c|c|}
\hline 
 $T_m(P),$ $^\circ$K & $v(P,T_m(P)),$ $\stackrel{\circ }{{\rm A}}$$^3$ & 
 $u_t,$ m/s & $G(P,T_m(P)),$ GPa & $G\,v/(k_BT_m)$  \\
\hline 
 205.59 & 35.698 & 952.6 & 1.686 & 21.21  \\
 190.90 & 36.216 & 909.7 & 1.516 & 20.84  \\ 
 175.91 & 36.785 & 879.5 & 1.395 & 21.14  \\
 162.80 & 37.319 & 843.0 & 1.263 & 20.98  \\
 162.07 & 37.350 & 847.0 & 1.274 & 21.28  \\
 148.19 & 37.959 & 800.0 & 1.118 & 20.75  \\
 134.47 & 38.601 & 768.6 & 1.015 & 21.11  \\
 123.16 & 39.155 & 736.0 & 0.918 & 21.15  \\
\hline 
  83.80 & 40.900 &       & 0.600 & 21.22  \\
\hline 
\end{tabular}
\end{center}
\vspace*{0.1cm}
Table 3. Numerical values of the ratio $G(P,T_m(P))v(P,T_m(P))/(k_BT_m(P))$ 
for Ar along its solidus. The last row of the table contains $P=0$ values. 
 \\
  
\subsection{Experimental verification}

{\it Direct} experimental validation of relation (1) over a restricted range 
of densities is possible for a single element, viz. argon. Ishizaki {\it et 
al.}~\cite{ISB} measured the transverse ultrasonic wave velocity, $u_t$, in 
compressed argon along its solidus as a function of temperature. We calculate 
the shear modulus from the formula $u_t=\sqrt{G/\rho },$ and $v=V/N_A$ from 
the measured argon melting curve \cite{CDC}, $V=V(T_m),$ $V$ being the molar 
volume. Our results for the values of $G\,v/(k_BT_m)$ are shown in Table 3.

For the $P>0$ data we find $G\,v/(k_BT_m)=21.06\pm 0.17,$ in agreement with 
its $P=0$ value (we get $G\,v/(k_BT_m)=21.08\pm 0.17$ for all of the data). 
Thus, $G\,v/(k_BT_m)$ for Ar deviates from a constant by less than 
1\%. 

\section{Model of the shear modulus at all temperatures and densities}

Preston and Wallace \cite{PW} constructed a model of the temperature 
dependence of the shear modulus ($0\leq T\leq T_m$) for arbitrary pressures. 
The $T$-dependence of $G$ involves two characteristic temperatures, namely 
the Debye temperature and the melting temperature. The shear modulus is 
always monotonically decreasing with $T,$ and is nonlinear for $T\stackrel{
<}{\sim }\Theta _D$ and linear from $\Theta _D$ to $T_m$ for most elements. 
An accurate representation of $G(T)$ at fixed density is achieved by ignoring 
the low-temperature non-linearity and approximating $G(T)$ as a linear function 
of the reduced temperature $T/T_m$ with the correct value $G(\rho ,0)$ at 
$T=0$ \cite{PW}: 
\beq 
G(\rho ,T)=G(\rho ,0)\left( 1-\beta \;\!\frac{T}{T_m(\rho )}\right) , 
\eeq 
In general, the parameter $\beta $ may be density dependent. A fit to 
shear-modulus data spanning temperatures from 
$T=0$ to $T/T_m\stackrel{>}{\sim }0.4$ at zero pressure gave $\beta _0=0.23\pm 
0.08$ \cite{PW}. (For the 11 fcc elements in Table 4 below $\beta _0=0.27\pm 
0.10.)$ On the other hand, $\beta ^{{\rm OCP}}=0.21\pm 0.18$ (as discussed 
above), which equals $\beta _0$ to within uncertainties, so we assume that 
$\beta $ is independent of density. At $\rho =\rho_{{\rm ref}}$ and $T=T_m(
\rho _{{\rm ref}}),$ Eq.\ (9) reduces to 
\beq
\beta =1-\frac{G(\rho _{{\rm ref}},T_m(\rho _{{\rm ref}}))}{G(\rho _{{\rm 
ref}},0)}.
\eeq

The linear temperature dependence is suggested by available $P=0$ experimental 
data on $G$ over the temperature range $0\leq T\leq T_m$ [1--19]. This 
straight-line representation turns out to be quite accurate: the maximum 
deviation of the data from the corresponding fitted lines is $\sim 5$ \% for 
21 of the 22 metals analyzed in \cite{PW}. The exception is uranium, for which 
$G(T)$ is nonlinear throughout the $\alpha $ phase at $P=0.$ As mentioned 
above, $G(T)$ is is nonlinear below $\Theta_D,$ thus $G(T)$ is nonlinear for 
low-melting-point solids from $T=0$ to $T_m.$ Despite the nonlinearity of 
$G(T)$ in these cases, the model uncertainty is only of order 10\%.

At any given pressure, the introduction of the temperature dependence of the 
density, $\rho =\rho (T),$ into Eq.\ (9) gives the temperature dependence 
of $G$ at that pressure. In Fig.\ 1 we compare $G(\rho (T),T)$ for 
$0\leq T\leq T_m$ at $P=0$ for Au and Cu to experimental data \cite{Au,Cu}. 
The temperature dependence of the density was taken from ref.\ \cite{Zin}, and 
$G(\rho,0)$ and $T_m(\rho )$ are described by Eqs.\ (13) and (14) below with 
parameter values from Tables 2 and 4. 

\vspace*{0.5cm}
\begin{center}
\epsfig{file=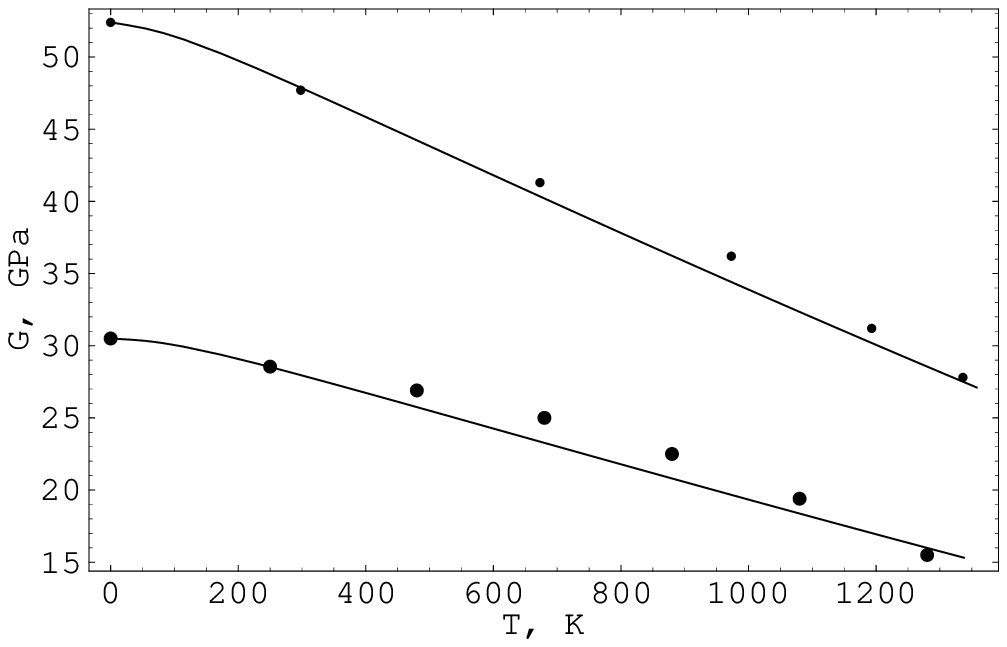,width=15cm,angle=0}
\end{center} 
Fig.\ 1. The $P=0$ shear moduli of Cu and Au: Eq.\ (9) with $\rho =
\rho (T)$ from ref.\ \cite{Zin} and $G(\rho (T),0)$ and $T_m(\rho (T))$ 
described by Eqs.\ (13) and (14) with the parameters from Tables 2 and 
4 vs.\ the experimental data on Cu \cite{Cu} (smaller points) and Au 
\cite{Au} (larger points). 
 \\

The Gr\"{u}neisen parameter was recently modeled as \cite{BP3} 
\beq
\gamma (\rho )=\frac{1}{2}\;\!+\;\!\frac{\gamma _1}{\rho ^{\;\!1/3}}\;\!+
\;\!\frac{\gamma _2}{\rho ^{\;\!q}},\;\;\;\gamma_1,\gamma _2,q={\rm const ,}
\;\;q>1, 
\eeq
through consideration of its low- and ultrahigh-pressure limits. This 
analytic form for $\gamma $ was obtained under the assumptions that 
(i) $\gamma \rightarrow 1/2$ as $\rho \rightarrow \infty ,$ (ii) $\gamma $ 
is an analytic function of $x\equiv 1/\rho ^{1/3},$ essentially the 
interatomic distance, and (iii) the coefficient of $x$ in the Taylor-Maclaurin 
series expansion for $\gamma $ is non-zero. The third term on the right-hand-
side of Eq.\ (11) represents the contribution of the quadratic and higher-order 
terms in the power series. The procedure for calculating the values of 
$\gamma _1$, $\gamma _2$, and $q$ is discussed below. 

Eq.\ (11) and the Lindemann criterion \cite{Young} 
\beq
\frac{d\ln T_m(\rho )}{d\ln \rho }=2\left( \gamma (\rho )-\frac{1}{
3}\right) 
\eeq
provide a model for the density dependence of the melting temperature, 
\beq
T_m(\rho )=T_m(\rho _{{\rm ref}})\left( \frac{\rho }{\rho _{{\rm ref}}}\right) 
^{1/3}\exp \left\{6\gamma _1\left( \frac{1}{(\rho _{{\rm ref}})^{\;\!1/3}}-
\frac{1}{\rho ^{\;\!1/3}}\right) +\frac{2\gamma _2}{q}\left( \frac{1}{(\rho _{
{\rm ref}})^{\;\!q}}-\frac{1}{\rho ^{\;\!q}}\right) \right\} .
\eeq
The natural choice for the reference density is $\rho _m$, the zero-pressure 
density at melt, which is known experimentally in most cases (see, e.g., 
\cite{Zin}). 

Finally, Eqs.\ (1),(9),(10), and (13) result in 
\beq
G(\rho ,0)=G(\rho _{{\rm ref}},0)\left( \frac{\rho }{\rho _{{\rm ref}}}\right) 
^{4/3}\exp \left\{ 6\gamma _1\left( \frac{1}{(\rho _{{\rm ref}})^{\;\!1/3}}-
\frac{1}{\rho ^{\;\!1/3}}\right) +\frac{2\gamma _2}{q}\left( \frac{1}{(\rho _{
{\rm ref}})^{\;\!q}}-\frac{1}{\rho ^{\;\!q}}\right) \right\} ,
\eeq  
where $\rho _{{\rm ref}}$ is most conveniently chosen to be either $\rho _m$ 
or $\rho _0,$ the density at zero pressure and temperature. 

Eqs.\ (9),(13), and (14) constitute our analytic model for the shear modulus. 
It requires the determination of 7 parameters, namely $\rho _{{\rm ref}},$ $G(
\rho _{{\rm ref}},0),$ $T_m(\rho _{{\rm ref}}),$ $\gamma _1,$ $\gamma _2,$ 
$q,$ and $\beta .$ The values of $\gamma _1,$ $\gamma _2$ and $q$ are obtained 
by simultaneous solution of Eq.\ (11) with $\rho =\rho \;\!(T=300\,^\circ $K) 
and $\rho =\rho _m,$ and Eq.\ (5) with $\Gamma _m=180$ \cite{BP3} and $T_m(
\rho )$ given by the high-density limit of Eq.\ (13). The value of $\gamma (
\rho _m)$ is obtained from the Kraut-Kennedy relation \cite{KK} and 
low-pressure melting data. The remaining parameters are either zero-pressure 
experimental data themselves or can be determined from such data (for example, 
$\beta )$. In Table 4 we present the values of $\rho _{{\rm ref}}$ (both 
$\rho _0$ and $\rho _m$), $G(\rho _0,0)$, $\gamma _1$, $\gamma _2$, $q$, and 
$\beta $ for all of the fcc elements of Table 2. The values of $G(\rho _m,0)$ 
can be calculated from the relation $G(\rho _m,0)=G(\rho _m,T_m)/(1-\beta )$ 
with $G(\rho _m,T_m)$ from Table 2, which also contains the values of $T_m(
\rho _m)$.  Since $\beta $-Co exists only above $T\approx 700\,^\circ $K 
at $P=0$, its values of $G(\rho _0,0)$ and $\beta $ were obtained from the 
conditions $G(\rho _m,T_m)=34.7$ and $G(\rho \;\!(T=710\,^\circ {\rm K})=8.62,
T=710\,^\circ {\rm K})=57.1$ \cite{Co}.

\vspace*{0.5cm}
\begin{center}
\begin{tabular}{|r|c|c|c|c|l|c|c|}
\hline 
 element & $\rho _0,$ g/cc & $\rho _m,$ g/cc & $G(\rho _0,0),$ GPa & 
 $\gamma _1,$ (g/cc)$^{1/3}$ & $\gamma _2,$ (g/cc)$^q$ & $q$ & $\beta $  \\
\hline 
         Ag & 10.63 & 9.850 & 33.5 & 2.23 & $9.63\cdot 10^4$    & 4.8 & 0.18 \\
         Al & 2.730 & 2.550 & 29.3 & 0.84 & 45.4                & 3.5 & 0.22 \\
         Ar & 1.771 & 1.622 & 1.46 & 1.06 & 6.42                & 2.2 & 0.23 \\
         Au & 19.49 & 18.29 & 30.5 & 3.21 & $1.97\cdot 10^{12}$ & 9.4 & 0.18 \\
$\beta $-Co & 8.910 & 8.180 & 73.2 & 1.81 & $6.28\cdot 10^4$    & 5.5 & 0.33 \\
         Cu & 9.020 & 8.370 & 52.4 & 1.87 & $2.31\cdot 10^4$    & 4.7 & 0.25 \\
         Ni & 8.970 & 8.220 & 93.6 & 1.85 & $5.60\cdot 10^5$    & 6.5 & 0.41 \\
         Pb & 11.60 & 11.05 & 11.7 & 3.09 & $8.21\cdot 10^8$    & 8.5 & 0.36 \\
         Pd & 12.13 & 11.29 & 50.3 & 2.40 & $3.34\cdot 10^6$    & 6.6 & 0.07 \\
         Pt & 21.58 & 20.19 & 66.3 & 3.21 & $1.13\cdot 10^{11}$ & 8.3 & 0.27 \\
         Rh & 12.49 & 11.49 & 158. & 2.16 & $1.46\cdot 10^7$    & 6.5 & 0.42 \\
\hline 
\end{tabular}
\end{center}
\vspace*{0.1cm}
Table 4. Numerical values of the model parameters for 11 fcc elements. The 
corresponding values of $T_m(\rho _m)$ and $G(\rho _m,T_m(\rho _m))$ are 
provided in Table 2.
 \\  \\


In Figs.\ 2, 3, and 4 we compare the melting curves of Ar, Al and Cu as given 
by Eq.~(13) with the corresponding parameters from Table 4 to experimental 
data. 

\vspace*{0.5cm}
\begin{center}
\epsfig{file=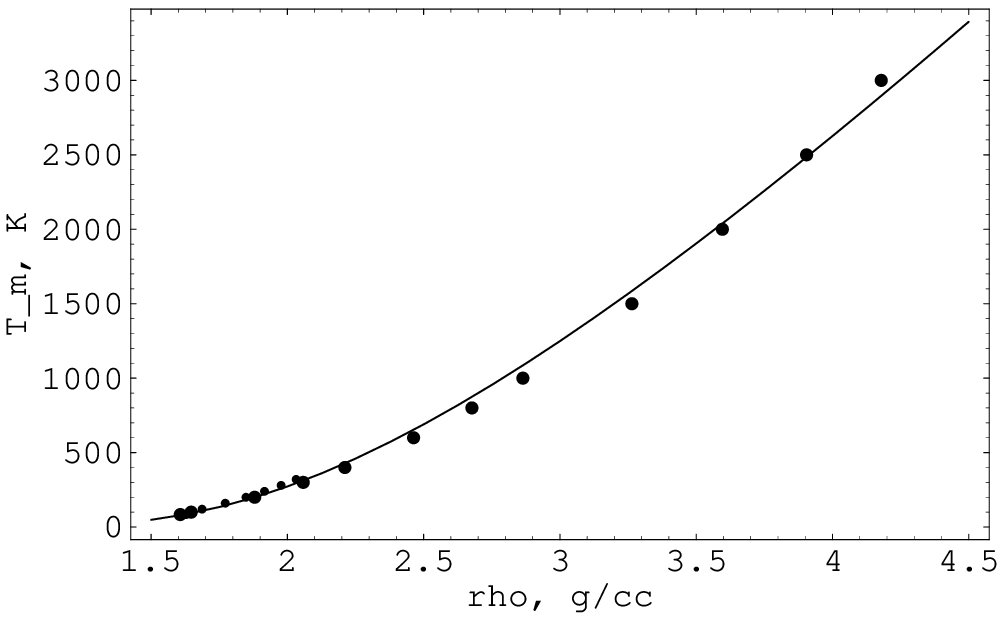,width=14cm,angle=0}
\end{center} 
Fig.\ 2. Melting curve of Ar: Eq.\ (13) with the Ar parameters from 
Table 4 vs.\ data. The smaller points are the experimental data of 
ref.\ \cite{CDC}, and the larger points are the results of calculations 
\cite{Ar-theor}. 
 \\

\vspace*{0.5cm}
\begin{center}
\epsfig{file=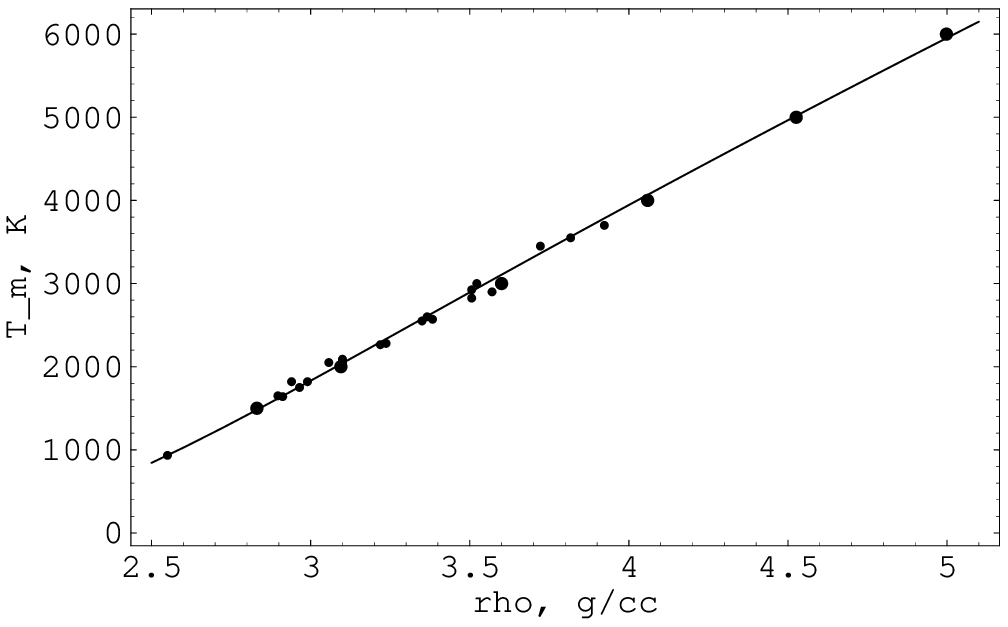,width=14cm,angle=0}
\end{center} 
Fig.\ 3. Melting curve of Al: Eq.\ (13) with the Al parameters from 
Table 4 vs.\ data. The smaller points are the experimental data of 
ref.\ \cite{Al-exp}, and the larger points are the results of calculations 
\cite{Al-theor}. 
 \\

\vspace*{0.5cm}
\begin{center}
\epsfig{file=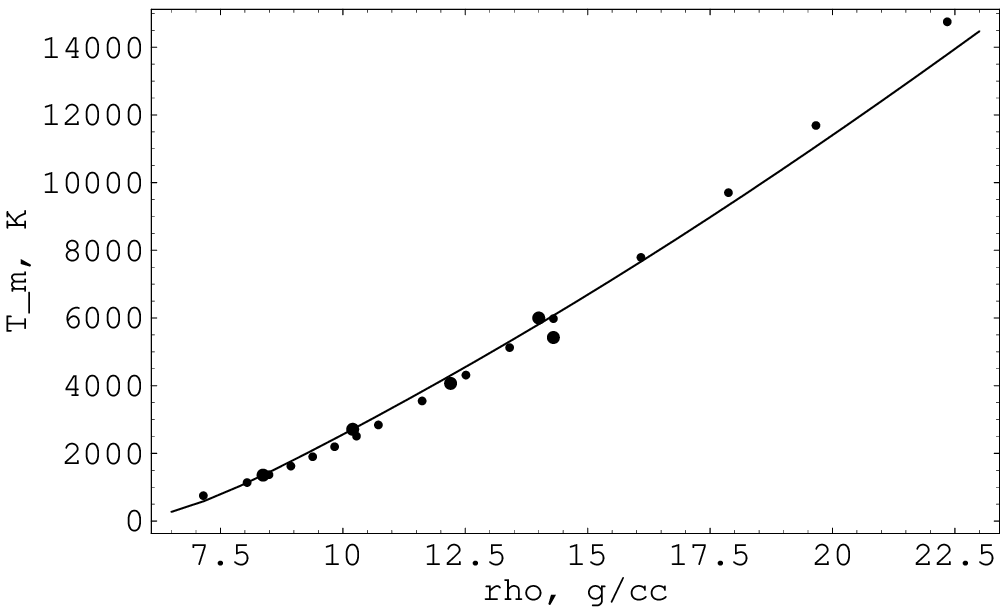,width=15cm,angle=0}
\end{center} 
Fig.\ 4. Melting curve of Cu: Eq.\ (13) with the Cu parameters from 
Table 4 vs.\ data. The smaller points are from a new SESAME melting curve 
table for Cu \cite{JD}. The larger points are the $P=0$ reference point at 
(in g/cc) $\rho =8.4,$ and the shock-melting points of ref.\ \cite{U} at 
$\rho =10.2,$ 12.2 and 14.3, and of refs.\ \cite{Mor,HHMcQ} at $\rho =14.0$.  
 \\  \\

Only five of the seven shear modulus parameters are independent because four 
appear in the model as two ratios, namely $\beta /T_m(\rho )$ in Eq.\ (9) and 
$G(\rho _{{\rm ref}},0)/(\rho _{{\rm ref}})^{\;\!4/3}$ in Eq.\ (14); hence 
the shear modulus is of the form 
\beq
G(\rho ,T)\;\!=\;\!a_1\;\rho ^{\;\!4/3}\;\!\exp \left\{ -\;\!\frac{a_2}{
\rho ^{\;\!q}}-\frac{a_3}{\rho ^{\;\!1/3}}\right\} \;\!-\;\!a_4\;\rho 
\;\!T,\;\;\;a_1,a_2,a_3,a_4,q={\rm const}>0. 
\eeq
As specific examples we provide the following formulas for 
the shear moduli of Ar, Al, Cu, and Au: 
\beq
G_{{\rm Ar}}(\rho ,T)\;\!=\;\!687.4\;\rho ^{\;\!4/3}\;\!\exp \left\{ 
-\;\!\frac{5.84}{\rho ^{\;\!2.2}}-\frac{6.36}{\rho ^{\;\!1/3}}\right\} 
\;\!-\;\!1.32\cdot 10^{-3}\;\!\rho \;\!T, 
\eeq
\beq
G_{{\rm Al}}(\rho ,T)\;\!=\;\!611.8\;\rho ^{\;\!4/3}\;\!\exp \left\{ 
-\;\!\frac{25.9}{\rho ^{\;\!3.5}}-\frac{5.04}{\rho ^{\;\!1/3}}\right\} 
\;\!-\;\!1.85\cdot 10^{-3}\;\!\rho \;\!T, 
\eeq
\beq
G_{{\rm Cu}}(\rho ,T)\;\!=\;\!841.2\;\rho ^{\;\!4/3}\;\!\exp \left\{ 
-\;\!\frac{9.83\cdot 10^3}{\rho ^{\;\!4.7}}-\frac{11.22}{\rho ^{\;\!
1/3}}\right\} \;\!-\;\!7.96\cdot 10^{-4}\;\!\rho \;\!T, 
\eeq
\beq
G_{{\rm Au}}(\rho ,T)\;\!=\;\!1022.0\;\rho ^{\;\!4/3}\;\!\exp \left\{ 
-\;\!\frac{4.19\cdot 10^{11}}{\rho ^{\;\!9.4}}-\frac{19.26}{\rho ^{\;\!
1/3}}\right\} \;\!-\;\!1.37\cdot 10^{-4}\;\!\rho \;\!T. 
\eeq 

\vspace*{0.5cm}
\begin{center}
\epsfig{file=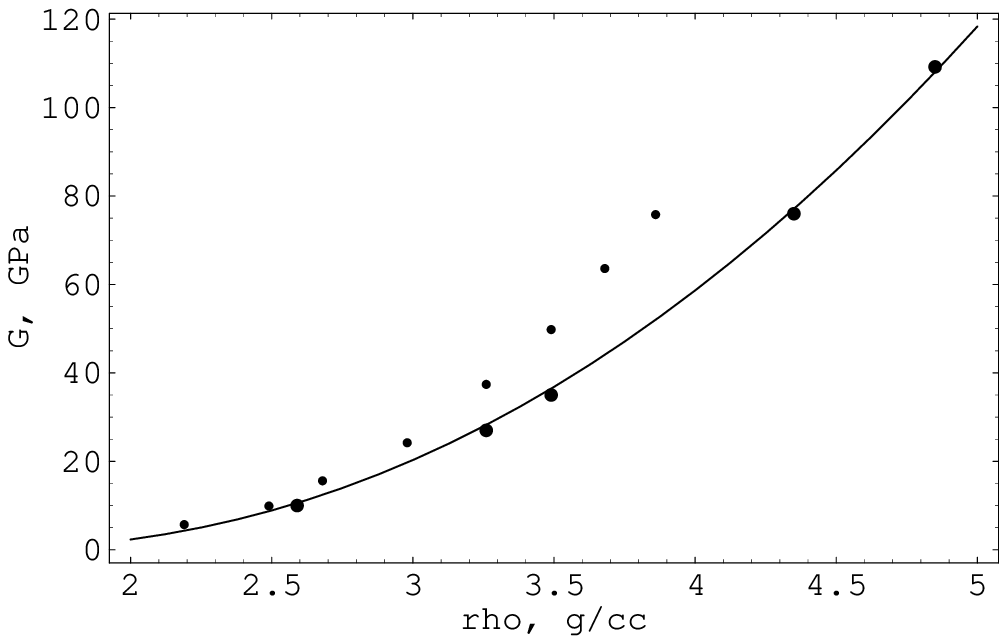,width=14cm,angle=0}
\end{center} 
Fig.\ 5. The $T=300$ shear modulus of Ar: Eq.\ (16) vs.\ older 
\cite{Ar-exp1} (smaller points) and more recent \cite{Ar-exp2} (larger 
points) experimental data. The experimental technique used to obtain 
the older data has been criticized \cite{Ar-exp2}. 
 \\

\vspace*{0.5cm}
\begin{center}
\epsfig{file=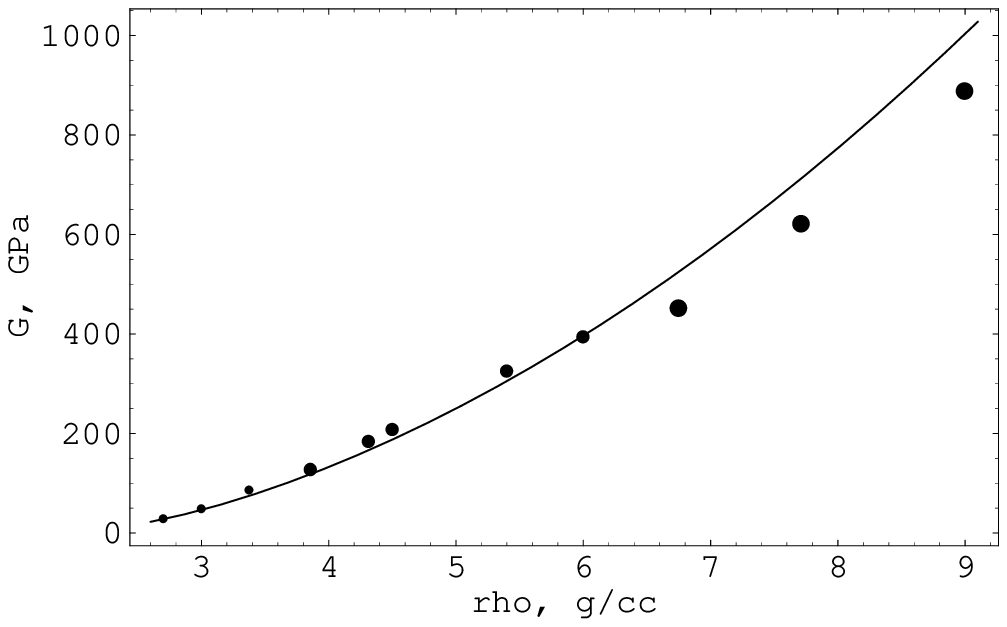,width=14cm,angle=0}
\end{center} 
Fig.\ 6. The $T=0$ shear modulus of Al: Eq.\ (17) vs.\ the 
electronic-structure calculations of ref.\ \cite{SS}. The small, medium, and 
large points represent the values of $G$ in fcc, hcp, and bcc phases of Al, 
respectively. 
 \\

In Fig.\ 5 we compare the $T=300\,^\circ $K shear modulus of Ar as given 
by Eq.\ (16) to experimental data. The $T=0$ shear modulus 
of Al from Eq.\ (17) is compared to the results of electronic structure 
calculations in Fig.\ 6. The $T=0$ shear moduli of Cu and Au as given, 
respectively, by Eqs.\ (18) and (19) are compared to the results of 
the corresponding electronic-structure calculations in Figs.\ 7 and 8 
in the next section. 

\section{Comparison of model to electronic-structure results for Cu and Au}

With the exception of Ar, experimental data are not available to test the 
model to megabar pressures. We can, however, test the $T=0$ version of the 
model by comparing it to the results of {\it ab initio} electronic-structure 
calculations of the shear modulus. 

Electronic structure calculations based on approximate density functional
theories have proven to give good predictions for a variety of material
properties. A study of the elastic constants of several elements 
and compounds \cite{Mehl} covering a wide range of elastic properties, 
found errors with respect to experiment of generally less than 10\% in the 
isotropic shear modulus. These results are obtained without empirical inputs. 
We expect such calculations to have similar accuracy under compression, thus 
providing a test of the new analytic model. 

For this reason, we have carried out electronic structure calculations to
obtain the single crystal elastic constants $C'=\left( C_{11}-C_{12}\right) 
/2,$ $C_{44},$ and $B=\frac{1}{3}\left( C_{11}+2C_{12}\right) $ for the fcc 
metals Cu and Au from normal to twice normal density. From these an average 
polycrystalline shear modulus is calculated and compared to the model. 

The method for the calculations was described by S\"{o}derlind {\it et 
al.}~\cite{Soderlind}. To evaluate $C'$, the lattice is deformed by the 
(volume conserving) transformation
\beq
\left( \begin{array}{ccc}
1+\delta  &    0      &         0        \\
0         & 1+\delta  &         0        \\
0         &    0      &  1/(1+\delta )^2 
\end{array} \right) .
\eeq
The resulting energy change is
\beq
\Delta E/V=6\;\!C'\;\!\delta ^2+O(\delta ^3).
\label{deltae_cprime}
\eeq
Similarly, $C_{44}$ is obtained by applying the (volume conserving) 
transformation
\beq
\left( \begin{array}{ccc}
    1   & \delta  &        0        \\
\delta  &     1   &        0        \\
    0   &     0   & 1/(1-\delta ^2)
\end{array} \right) ,
\eeq
resulting in an energy change
\beq
\Delta E/V=2\;\!C_{44}\;\!\delta^2+O(\delta ^4).
\label{deltae_c44}
\eeq 
In our calculations we have evaluated the energy as a function of $\delta $
at intervals of 0.01, up to $\delta =0.04$. For the $C'$ case the energy is
not an even function of $\delta $, and so negative values of delta were used.
The resulting $E(\delta )$ were fit to 4$^{{\rm th}}$ degree polynomials
and the quadratic coefficent was used to evaluate the elastic constant from
Eq.~(\ref{deltae_cprime}) or Eq.~(\ref{deltae_c44}). 

The bulk modulus $B$ is obtained from the volume-dependent energy of the 
undistorted crystal by
\beq
B=V\;\!\frac{d^2E}{dV^2}.
\eeq
The energy was evaluated at volume intervals of 5\% of the normal volume,
from 20\% expanded to 50\% contracted. Derivatives were evaluated by fitting 
the equation of state of Rose {\it et al.} \cite{Rose} to the energies and
differentiating the function. It was found that a single curve of this
type did not accuaretly fit both the high density points and the points
near the minimum, so seperate overlapping fits were made for the 10 highest
and lowest densities.

The electronic structure calculations were based on the linearized augmented 
plane-wave (LAPW) code WIEN97 \cite{Wien97}. The energy functional used was 
the generalized-gradient approximation as parameterized by Perdew, Burke, 
and Ernzerhof \cite{Perdew}. Some numerical parameters used in the 
calculations for Cu (Au) were, in atomic units:
muffin tin radius $r_{\rm MT}=1.8$ (2.0), plane wave 
cut-off $r_{\rm MT}k_{\rm max}=9.0$, cut-off for expansion of 
density and potential $g_{\rm max}=16$ (19); Brillouin zone integrals
used special points corresponding to $16^3$ ($18^3$) points in the
full zone, with Gaussian smearing of the energies by 20 mRy.

Our results on $C',$ $C_{44}$ and $B$ for Cu and Au are shown in Tables 5 and 
6, respectively. It is interesting to note the increasing anisotropy of Au 
under pressure. From Table 6 we see that, for Au, $C'$ does not increase 
nearly as rapidly as $C_{44}$ with compression. This is connected with the 
fact that the energy difference between the fcc and bcc structures is small 
at all pressures \cite{Boettger}. The distortion corresponding to $C'$ 
is along the Bain path connecting fcc to bcc, and it has been seen 
\cite{Soderlind} that a small energy difference between these structures 
correlates with a small value of $C'$.

Let us now turn to the calculation of the shear moduli of Cu and Au. 
For a solid of cubic crystalline structure, as analysis by Kr\"{o}ner 
\cite{Kroner} shows, successively narrower bounds can be placed on the 
shear modulus as the degree of disorder in grain orientation increases. 
In the limit of perfect disorder, the value of the shear modulus is 
the single positive real root of a cubic equation with coefficients that 
depend on the single-crystal elastic constants $C',$ $C_{44},$ and $B:$ 
\beq
x^3\,+\,\frac{9B+4C'}{8}\,x^2\,-\,\frac{3\left( B+4C'\right) C_{44}}{8}\,x\,
-\,\frac{3BC'C_{44}}{4}\,=\,0.
\label{G}
\eeq

The values of the shear modulus calculated from Eq.~(\ref{G}) are shown in 
Tables 5 and 6 along with $C',$ $C_{44}$ and $B.$ 

As a by-product of our analysis, we obtain the interesting results that 
$G(2\rho _0,0)\simeq 10\,G(\rho_0,0)$ for Cu, and 
$G(2\rho _0,0)\simeq 20\,G(\rho_0,0)$ for Au.

In Figs.\ 7 and 8 we compare Eqs.\ (18) and (19) with $T=0,$ for Cu and Au, 
to the corresponding $G$ entries in Tables 5 and 6. 

\vspace*{0.5cm}
\begin{center}
\begin{tabular}{|c|c|c|c|c|}
\hline 
 $\rho ,$ g/cc & $C',$ GPa & $C_{44},$ GPa & $B,$ GPa & $G,$ GPa   \\
\hline 
 8.850 & 30.404 & 77.639 & 142.15 & 53.912   \\
 9.833 & 41.863 & 124.78 & 235.73 & 81.901   \\ 
 11.06 & 48.599 & 167.42 & 386.02 & 104.66   \\
 12.64 & 83.020 & 260.81 & 652.48 & 168.57   \\
 14.75 & 130.48 & 445.26 & 1151.8 & 279.70   \\
 17.70 & 229.71 & 800.23 & 2118.4 & 499.25   \\
\hline 
\end{tabular}
\end{center}
\vspace*{0.1cm}
Table 5. The single-crystal elastic constants and shear modulus of Cu 
as functions of density from the electronic-structure calculations 
described in the text. 
 \\

\begin{center}
\begin{tabular}{|c|c|c|c|c|}
\hline 
 $\rho ,$ g/cc & $C',$ GPa & $C_{44},$ GPa & $B,$ GPa & $G,$ GPa   \\
\hline 
 19.29 & 16.445 & 31.690 & 201.20 & 24.585   \\
 21.43 & 19.550 & 77.940 & 339.57 & 46.764   \\ 
 24.11 & 33.890 & 127.75 & 568.39 & 78.093   \\
 27.56 & 35.053 & 255.22 & 1029.0 & 127.48   \\
 32.15 & 69.837 & 479.27 & 1918.2 & 243.34   \\
 38.58 & 121.16 & 912.15 & 3753.2 & 451.65   \\
\hline 
\end{tabular}
\end{center}
\vspace*{0.1cm}
Table 6. The single-crystal elastic constants and shear modulus of Au 
as functions of density from the electronic-structure calculations 
described in the text. 
 \\
 
\begin{center}
\epsfig{file=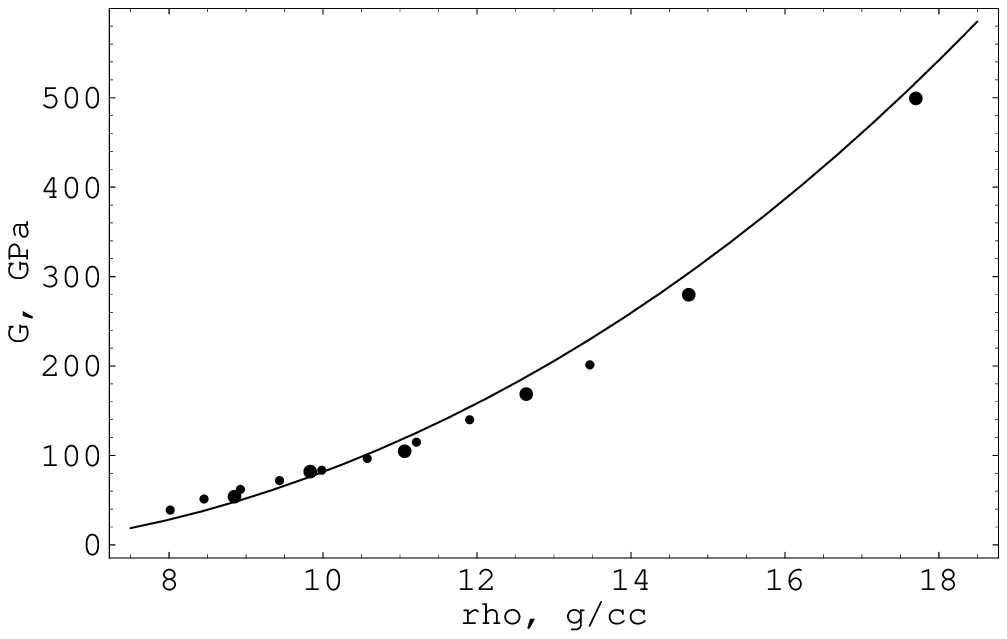,width=14cm,angle=0}
\end{center} 
Fig.\ 7. The $T=0$ shear modulus of Cu: Eq.\ (18) vs.\ 
electronic-structure calculations (larger points, Table 5). 
The smaller points, obtained from first-principles 
calculations \cite{RJGA}, are shown for comparison. 
 \\

\vspace*{0.5cm}
\begin{center}
\epsfig{file=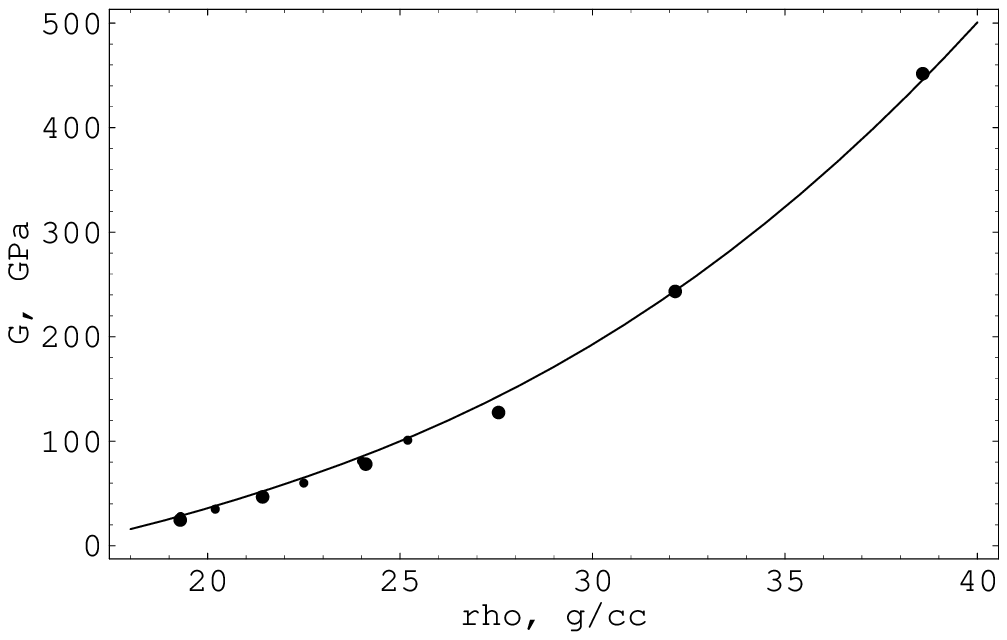,width=14cm,angle=0}
\end{center} 
Fig.\ 8. The $T=0$ shear modulus of Au: Eq.\ (19) vs.\ 
electronic-structure calculations (larger points, Table 6). 
The smaller points, obtained from first-principles 
calculations \cite{Au-exp}, are shown for comparison. 
 \\

\vspace*{0.5cm}
\begin{center}
\epsfig{file=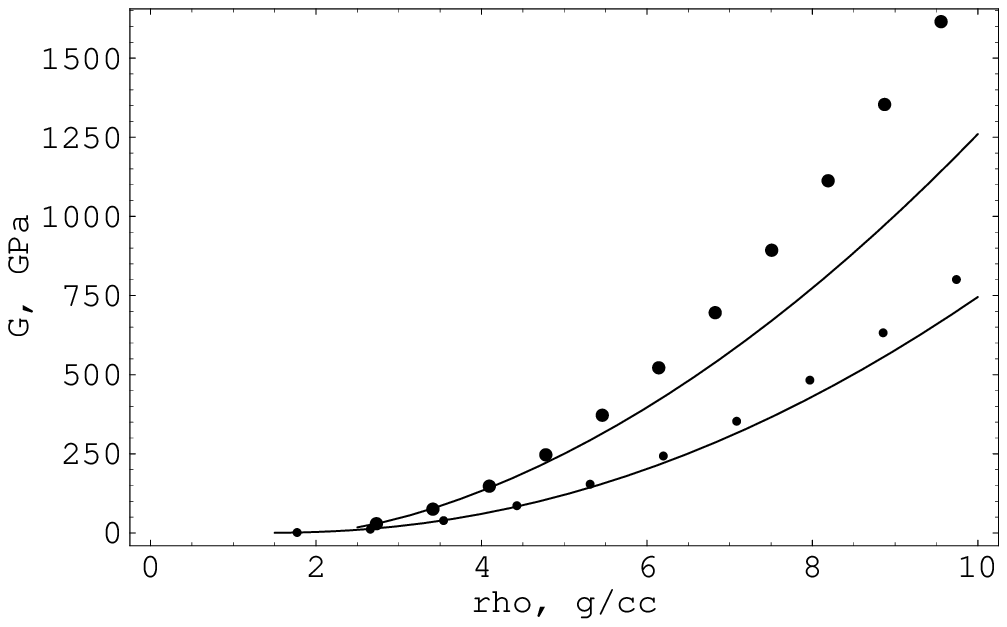,width=14cm,angle=0}
\end{center} 
Fig.\ 9. Comparison of the two models for the $T=0$ shear modulus: 
Eqs.\ (16) and (17) vs.\ the corresponding Guinan-Steinberg values 
for Ar (smaller points) and Al (larger points). 
 \\

Finally, it is interesting to compare our model at $T=0$ to the 
Guinan-Steinberg model mentioned in the Introduction. The equation of state, 
$P=P(\rho )$, is needed to make this comparison. In Fig.\ 9 the models are 
compared to each other for Ar and Al. The corresponding equations of state 
are taken from ref.\ \cite{HS}. For Al, $G_0'=1.8$ comes from ref.\ \cite{GS2}. 
For Ar, $G_0'=1.6$ is obtained from the relation \cite{BPS2} $\gamma _0=B_0/
2\,G_0'/G_0-1/6$ with $\gamma _0=\gamma (\rho _0)$ from Eq.\ (11) and $B_0$ 
taken from ref.\ \cite{HS}. The values of $\rho _0$ and $G_0=G(\rho _0,0)$ 
can be found in Table 4. 

It is seen that agreement between the two models is good at low densities, 
but it gradually deteriorates with increasing compression. The reason for 
this must be that the Guinan-Steinberg model generally provides only the 
correct functional form $G\sim \rho ^{\;\!4/3}$ in the limit of infinite 
compression, not the precise numerical value of $G$ in that limit, in 
contrast to our model which provides both. 

\section{Concluding remarks}

We have constructed an analytic model of the shear modulus applicable at all 
densities greater than or of order ambient ($G(\rho ,0)\rightarrow 0$ as 
$\rho \rightarrow 0,$ as required, but the model may not be quantitatively 
correct for expanded states), and temperatures from 0 to $T_m$. All of the 
model parameters can be obtained from low-pressure experimental data. The 
model has the proper low-pressure 
and high-pressure limits, by construction, and to within uncertainties it 
agrees with electronic-structure values of $G$ for Cu and Au to compressions of 
2, which roughly corresponds to pressures of 5 Mbar for Cu and 9 Mbar for Au.

The above comparisons of our shear modulus model, which includes a model 
for $T_m(\rho ),$ to electronic structure calculations and experimental data
on Ar, Al, Cu, and Au show very good agreement. This suggests that our model
accurately represents the density and temperature dependence of the shear 
moduli of monatomic solids in general. There is, however, no theoretical 
justification for applying our model to alloys or compounds, although in 
practice it may work reasonably well in these cases. Its generalization 
to more complex materials would involve generalizing our model for the 
Gr\"{u}neisen parameter. A functional form for $\gamma (\rho )$ depends 
critically on the asymptotic $(\rho \rightarrow \infty )$ form of the 
equation of state \cite{BP3}, and it has been suggested that the asymptotic 
forms of the equations of state of more complex materials, e.g., ionic, 
covalent, or molecular crystals, are different from that of a metal\cite{Dav}. 
If so, the limiting value of $\gamma $ is unknown (not necessarily 1/2) for 
such materials. In that case, an analytic model for the Gr\"{u}neisen 
parameter cannot be constructed, hence analytic forms for the melting 
curve and shear modulus cannot be obtained.


We now briefly discuss three potential applications of our model.

(1) Plastic deformation of metals at high pressure. It is generally assumed 
that the ratio of the plastic flow stress (shear stress necessary to induce 
plastic deformation at a given strain rate) to the shear modulus is 
approximately independent of pressure.  In other words, the predominant 
pressure dependence of the plastic flow stress is contained in the shear 
modulus. An accurate, simple analytic (for fast evaluation) model of the 
shear modulus is therefore essential for numerical simulations of material 
deformation over extremes in pressure.

(2) Numerical simulations of elastic wave propagation, including pressure 
release waves in shocked solids. The differential stress deviator, $ds_{ij}$, 
is equal to $2G(\rho, T)\,(d\epsilon_{ij}-\delta_{ij}\,d\epsilon_{kk}/3)$ 
plus material rotation terms ($d\epsilon_{ij}$ is the differential elastic 
strain), thus a model of the shear modulus is required for calculations of 
elastic wave propagation in materials with sufficiently high yield stresses 
that the stress deviators are not negligible. The speed of a release wave 
in a shocked solid depends on $G(\rho_H,T_H),$ where $\rho _H$ and $T_H$ 
are the density and temperature of the shocked state. 
 
(3) Pulsations and quakes of dense stars. Hansen and Van Horn \cite{HVH} 
have done a preliminary analysis of the effects of crystalline cores on 
the oscillations of white dwarfs and found that the $g$-like spheroidal mode 
frequencies are increased by approximately a factor of two, concluding that 
the elastic shear strength of the core must be taken into account in the 
computation of cool white dwarf oscillations. The inclusion of elastic 
shear strength in the neutron star pulsation equations of McDermott {\it et 
al.}~\cite{MVHH} resulted in the appearance of two classes of oscillation 
modes not present in a fluid neutron star. The change in the shape of the 
surface following a neutron star quake is proportional to the shear 
modulus of the crust \cite{Rud}. 

Further tests of our model for the shear modulus should be made as 
high-pressure experimental data and electronic structure results become 
available for elements other than argon, aluminum, copper and gold. New 
zero-pressure data are also needed to generate additional sets of model 
parameters. 

\section{Acknowledgements} 
We wish to thank J.C. Boettger, J.D. Johnson and G.W. Pfeufer for very 
stimulating discussions on the subject of the shear modulus. 


\end{document}